\documentclass[usenatbib]{mn2e}

\pdfoutput=1

\usepackage{graphicx}
\usepackage[usenames]{color}
\usepackage{amssymb}

\newlength{\colwidth}
\newcommand{\Mpch}{h^{-1}\,\mbox{Mpc}}

\newcommand{\Mth}{M_{200}}
\newcommand{\Rth}{R_{200}}
\newcommand{\cth}{c_{200}}
\newcommand{\Vth}{V_{200}}

\newcommand{\age}{z_{0.5}}
\newcommand{\relp}{r}
\newcommand{\ar}{s}
\newcommand{\TT}{T}

\newcommand{\MR}{f_{\rm sub}}
\newcommand{\Msub}{M_{\rm sub}}

\newcommand{\dtwo}{D_{1,0.1}}
\newcommand{\rank}{R_{\rm S}}

\def\mathbi#1{\textbf{\em #1}}
\begin{document}

\title[Correlations of dark matter halo properties]{The correlation structure of dark matter halo properties}

\author[A. Jeeson-Daniel et al.]{Akila Jeeson-Daniel$^{1,2}$\thanks{E-mail: akila@mpa-garching.mpg.de}, Claudio Dalla Vecchia$^{2,3}$, Marcel R. Haas$^{2,4}$, and Joop Schaye$^2$\\
$^1$Max Planck Institute for Astrophysics, Karl-Schwarzschild Stra\ss{}e 1, 85741 Garching, Germany\\
$^2$Leiden Observatory, Leiden University, P.O. Box 9513, 2300 RA Leiden, the Netherlands\\
$^3$Max Plank Institute for Extraterrestrial Physics, Gissenbachstra\ss{}e 1, 85748 Garching, Germany\\
$^4$Space Telescope Science Institute, 3700 San Martin Drive, Baltimore, MD 21218, USA}

\maketitle

\abstract We investigate the correlation between nine different dark matter halo properties using a rank correlation analysis and a Principal Component Analysis for a sample of haloes spanning five orders of magnitude in mass. We consider mass and dimensionless measures of concentration, age, relaxedness, sphericity, triaxiality, substructure, spin, and environment, where the latter is defined in a way that makes it insensitive to mass. We find that concentration is the most fundamental property. Except for environment, all parameters are strongly correlated with concentration. Concentration, age, substructure, mass, sphericity and relaxedness can be considered a single family of parameters, albeit with substantial scatter. In contrast, spin, environment, and triaxiality are more independent, although spin does correlate strongly with substructure and both spin and triaxiality correlate substantially with concentration. Although mass sets the scale of a halo, all other properties are more sensitive to concentration.
\endabstract


\section{Introduction}

Hierarchical structure formation in the $\Lambda$CDM universe predicts
that galaxies form within virialised dark matter (DM)
haloes which merge to form larger ones. Understanding the structure of dark matter haloes is therefore of crucial importance for models of the formation of galaxies. A large number of 
simulations have been used to study the correlations between halo
properties like mass, concentration, accretion history, spin, shape,
substructure and environment
\citep[e.g.][]{Navarro1997, Bullock2001, Harker2006, Bett2007, Maccio2007, Hahn2007, Duffy2008, Wang2010, Haas2011} and to investigate the dependence of halo clustering on several of these
parameters \citep[e.g.][]{Gao2005, Wechsler2006, Bett2007, Gao2007,
Jing2007, Wetzel2007, Li2008}. Semi-analytic and halo occupation models usually assume halo mass to be the most important or even the only property that defines the halo and its galaxy content
\citep[e.g.][]{Mo1996,Cooray2002}. However, there is mounting evidence that formation time is likely to be another very important parameter \citep{Croton2007,Taylor2011}. 

In this work we present a numerical study of the relations between halo mass, formation history, shape, dynamical state, and environment at redshift $z=0$. We make use of both a correlation and a Principal Component Analysis (PCA) for a set of 9 parameters. By combining 5 large, cosmological N-body simulations, we are able to explore a mass range that spans 5 orders of magnitude with no particular mass dominating the distribution.

This paper is organised as follows. In Section~\ref{sec:properties} we define the halo properties that we consider and we discuss the simulations in Section~\ref{sec:simulations}. In Section~\ref{sec:analysis_and_results} we discuss the analysis methods and present the results of our investigation. 

\section{Definitions of halo properties}
\label{sec:properties}

We will consider the 9 halo properties described below. Except for mass, all are dimensionless.

\begin{description}
\item[\textbf{Virial mass,} $\mathbi{M}_{\mathbf{200}}$]
The virial mass is the mass within a sphere of radius $\Rth$
centered on the most bound particle of the halo, where $\Rth$ is the
radius within which the overdensity is $200$ times the critical
density of the Universe.

\item[\textbf{Concentration,} $\mathbi{c}_{\mathbf{200}}$]
The concentration of a halo is found by fitting an NFW
density profile \citep{Navarro1997} to the particle distribution and
calculating $\cth \equiv\Rth/r_{\rm s}$ where $r_{\rm s}$ is the NFW scale
radius. The values were obtained from \cite{Duffy2008}.

\item[\textbf{Age,} $\mathbi{z}_{\mathbf{0.5}}$]
The age of a halo is defined as the redshift at which at least half of the final mass, i.e. $0.5\,\Mth$, is in FOF haloes with at
least one-tenth of the final mass. The redshift and mass
fraction are calculated by linear interpolation between adjacent
snapshots. 

\item[\textbf{Relaxedness,} $\mathbi{r}$]
The relaxedness parameter is defined as the distance between the most bound particle and the centre of mass of the FOF halo divided by $\Rth$. A more relaxed halo has a smaller $\relp$.

\item[\textbf{Sphericity,} $\mathbi{s}$]
The halo sphericity, $\ar\equiv c/a$, is defined as the ratio of the length of the minor axis $c$ to the length of the major axis $a$. The axis lengths are calculated by diagonalising the inertia tensor defined as in \cite{Bett2007}. 

\item[\textbf{Triaxiality,} $\mathbi{T}$]
Triaxiality, $\TT \equiv (a^2-b^2)/(a^2-c^2)$, where $b$ is the length of the intermediate axis of the diagonalised inertia tensor. $\TT\rightarrow 1$ for a highly prolate halo and $\TT\rightarrow 0$ for a highly oblate halo.

\item[\textbf{Substructure,} $\mathbi{f}_{\mathbf{sub}}$]
The substructure parameter, $\MR\equiv \Msub/\Mth$, is defined as the fraction of the mass within $\Rth$ that is in substructures. We consider only sub-haloes with maximum circular velocities larger than $10^{-0.5}$ times that of the parent halo.

\item[\textbf{Spin,} $\mathbf{\lambda}$]
We use the modified spin parameter, $\lambda\equiv j (\sqrt{2}\Vth\Rth)^{-1}$, where $j$ is the specific angular momentum and $V_{200}$ the circular velocity of the halo \citep{Bullock2001}.  For details and a comparison of different spin parameter definitions, see \citet{Maccio2007}.

\item[\textbf{Environment,} $\mathbi{D}_{\mathbf{1,0.1}}$]
The environment in which the halo is residing is defined as
the distance to the nearest FOF halo with mass greater
than $0.1\,\Mth$, divided by the neighbour's $\Rth$. This measure of environment is related to the inverse of the tidal force exerted by the neighbour and, contrary to other environmental parameters used in the literature, does not correlate with halo mass \citep{Haas2011}.

\end{description}

\section{Simulations}
\label{sec:simulations}

We use a sample of five cosmological N-body simulations with box sizes increasing by factors of two from $L=25~\Mpch$ to $400~\Mpch$. The
simulations were run with the TreePM code \textsc{gadget-3}
\citep{Springel2005}. We adopted a set of cosmological parameters consistent with the 7-year WMAP data \citep{Komatsu2011}: $(\Omega_{\rm m},\Omega_{\Lambda},h,\sigma_8,n_{\rm s}) = (0.258,0.742,0.719,0.796,0.963)$.
Glass-like cosmological initial conditions were generated
at redshift $z = 127$ using the Zeldovich approximation
and a transfer function generated with \textsc{cmbfast} \citep[v.~4.1]{Seljak1996}. Each simulation contains $512^3$ collisionless particles. The particle mass is $1.16\times 10^7~\mbox{M}_{\odot} (L/25~\Mpch)^3$. The gravitational softening was set to $1/25$th of the mean inter-particle spacing, but was held fixed below $z=3$. 

Catalogues of dark matter haloes were produced using a friends-of-friends (FOF) algorithm with dimensionless linking length $b=0.2$. The FOF haloes were further analysed using the \textsc{subfind} substructure finder \citep{Springel2001}.

Different parameters require different resolutions, with $\MR$ and $\cth$ being the most demanding. To ensure complete convergence for all properties, we only use haloes with at least $10^4$ particles. We verified that using all haloes with at least $10^3$ particles gives nearly identical results. The
selected sample includes 1867 haloes (488, 534, 492, 306
and 47, respectively from the $L=25$, 50, 100, 200 and
$400~\Mpch$ simulations).

\begin{figure*}
\includegraphics[width=0.95\textwidth]{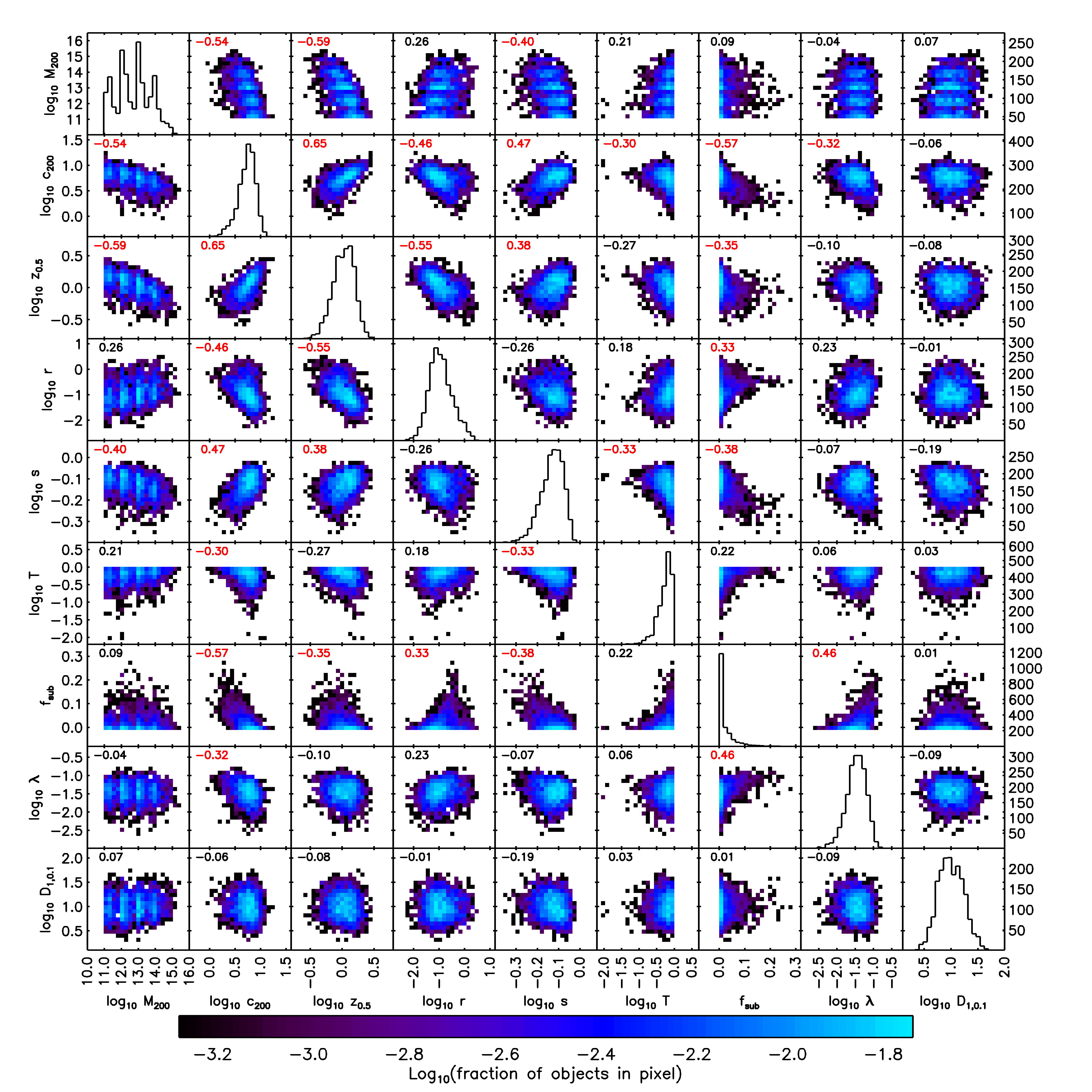}
\caption{Correlations between different halo properties. The colour scale indicates the logarithm (base 10) of the fraction of haloes in the pixel. The Spearman rank correlation coefficient is shown in the top left of each panel. To guide the eye, values $\left |R_S\right |\ge 0.3$ are printed red. The panels on the diagonal show histograms of the parameter values with the right vertical axes showing the number of haloes in the bins. Note that the figure is symmetric with respect to the diagonal.}
\label{ccdp1}
\end{figure*}

\begin{figure*}
\includegraphics[width=0.95\textwidth]{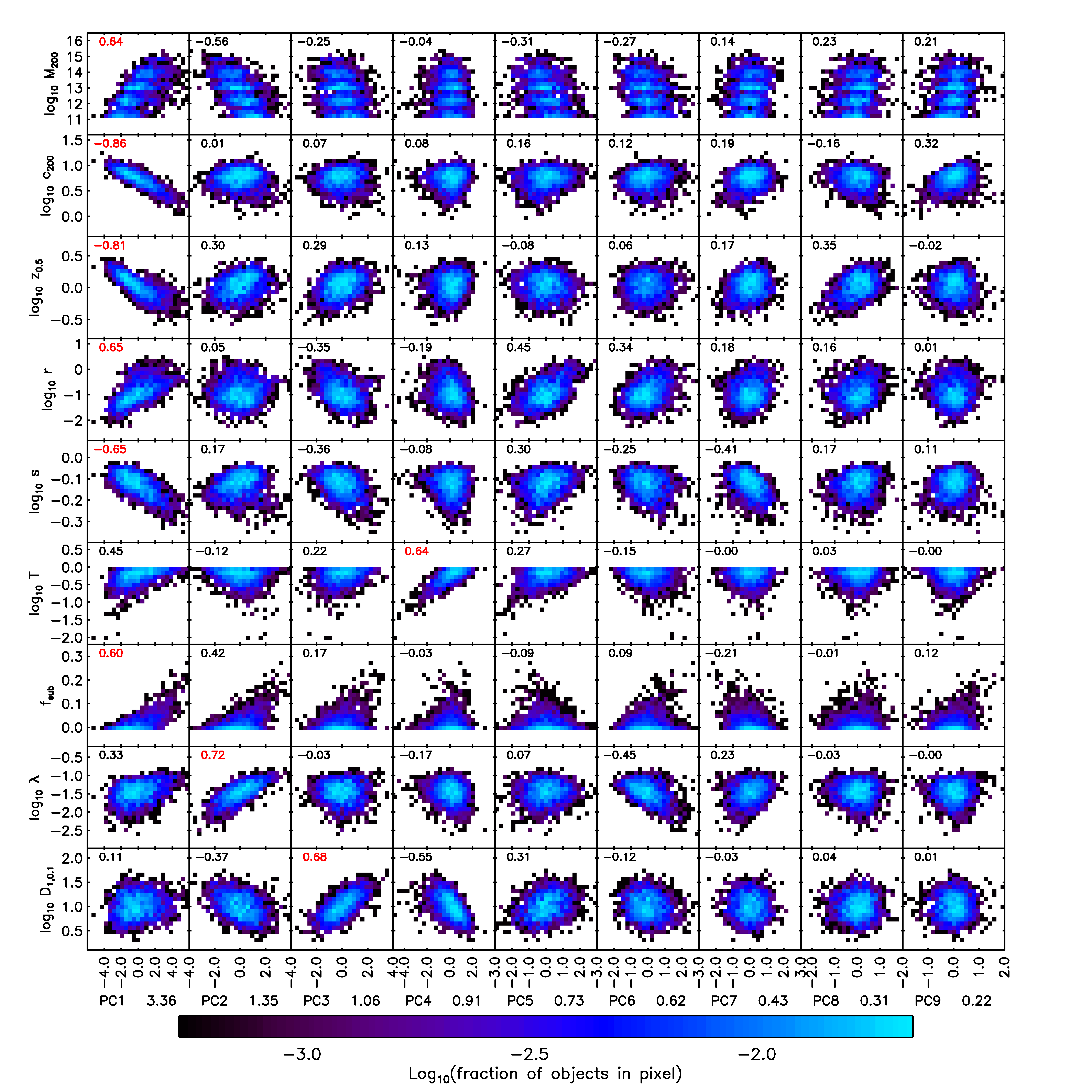}
\caption{Correlations between different halo properties and principal components. The colour scale indicates the logarithm (base 10) of the fraction of haloes in the pixel. The Spearman rank correlation coefficient is printed in the top left of each panel. For each parameter, the value of the strongest correlation is printed in red. The eigenvalue of each PC is shown along the horizontal axis.}
\label{pcdp1}
\end{figure*}

\section{Results}
\label{sec:analysis_and_results}
The relations between the different parameters and their
relative importance are investigated using a non-parametric correlation analysis and a principal component analysis in \S\ref{sec:correlation} and \S\ref{sec:pca}, respectively.

\subsection{Correlation analysis}
\label{sec:correlation}
Figure~\ref{ccdp1} shows how the 9 halo properties correlate with each other. Each panel shows the distribution of haloes projected onto the 2 dimensions that are plotted along the axes. The colour scale indicates the logarithm of
the fraction of haloes in a pixel. The value of the Spearman rank
correlation coefficient, $\rank$, is printed in each panel. Values of $\rank$ run from $-1$ (perfect anti-correlation) to $1$ (perfect
correlation), while $\rank = 0$ means \textit{no} correlation. For reference, for our sample of 1867 haloes, values $\left |\rank\right |\ge 0.06$ reject the hypothesis that no correlation is present at greater than 99\% confidence. For the panels along the diagonal the two dimensions are identical and we plot instead histograms of the values for the corresponding parameter with the right vertical axes showing the number of haloes per bin. Note that the figure is symmetric with respect to the diagonal.

The strongest correlation, with $\rank=0.65$, is between age ($\age$) and concentration ($\cth$), with older haloes being more concentrated. This correlation is expected, as the concentration is thought to reflect the time since the assembly of the inner part of the halo \citep[e.g.][]{Zhao2003}. Substructure ($\MR$), mass ($\Mth$), sphericity ($\ar$), and relaxedness ($\relp$) are all also strongly correlated ($|\rank| \ge 0.46$) with concentration. More concentrated haloes tend to have less substructure, be less massive, more spherical, and more relaxed. This can also be understood, as more concentrated haloes are older and have thus had more time to relax and to tidally strip substructures. Moreover, tidal forces, and thus the rate at which substructures are destroyed, will be larger in more concentrated haloes. More concentrated haloes also tend to have lower spins ($\lambda$) and to be more oblate (smaller $\TT$), but these correlations are less strong ($\rank \approx 0.3$). Age shows nearly the same correlations as concentration, although it is interesting to note the anti-correlations with substructure and spin are much weaker for age than for concentration. 

The two strongest correlations that involve neither concentration nor age are between substructure and spin ($\rank=0.46$) and between sphericity and mass ($\rank=-0.40$). The former correlation may reflect the fact that orbital motions of satellites contribute to the spin. Triaxiality and environment appear to be the most independent parameters, correlating most strongly with sphericity ($\rank = -0.33$ for triaxiality and -0.19 for environment). 

Note that our measure of environment was designed to be insensitive to mass, which it indeed is ($\rank = 0.07$). It is thus quite different in nature from the environmental parameters used in the literature, which all correlate strongly with mass \citep{Haas2011}. The parameter $\dtwo$ measures the distance to the nearest halo that is at least 0.1 times as massive and expresses this distance in units of the neighbour's virial radius. Typical values are $\sim 10$ which, given that most neighbours have masses near the lowest allowed values, implies that the nearest neighbours are typically at distances of a few virial radii. The absence of strong correlations with $\dtwo$ may indicate that all aspects of the environment that do influence the properties of haloes are correlated with halo mass. 

It is interesting that all halo properties correlate more strongly with concentration than with mass (except for environment, which, however, does not correlate much with either parameter). The difference is particularly striking for substructure and spin. While we find no evidence for a dependence of spin on mass ($\rank = -0.04$), spin is substantially anti-correlated with concentration ($\rank = -0.32$), as also found by \citet{Maccio2007}. In agreement with \citet{Gao2004}, substructure is nearly independent of mass ($\rank=0.09$), but strongly anti-correlated with concentration ($\rank=-0.57$). It thus appears that concentration, and the very closely related parameter age, may in some sense be more fundamental than mass. A better way to judge which (combinations of) parameters are most fundamental, is to carry out a PCA, which we will do next. 

\subsection{Principal Component Analysis}
\label{sec:pca}
PCA is a statistical technique to find the number of independent
parameters which are needed to account for the variance in a data
set.  Most of the variance in the data set can be represented by a
subset of all eigenvectors of the covariance matrix. These
eigenvectors are known as the principal components (i.e., PC$1$,
PC$2$, etc.). The eigenvalues corresponding to the eigenvectors
indicate whether or not that PC is important. If the variables are standardized (i.e.\ transformed to have mean zero and variance one), then the fraction of the total variance in the data set due to a given PC is equal to its eigenvalue divided by the total number of parameters. 

Except for $\MR$, which can be zero, we use the logarithm of the parameters. Before running the PCA, we standardize the halo properties by subtracting the means and dividing by the standard deviations. The results are insensitive to whether we take the logarithm or not, but standardizing is crucial because otherwise the result would depend on the choice of mass unit (note that all other parameters are dimensionless). 

Figure~\ref{pcdp1} shows the results of the PCA. Each panel shows the correlation between the halo property plotted along the vertical axis and the PC plotted along the horizontal axis. As in Figure~\ref{ccdp1}, the values of the Spearman rank correlation coefficients are shown and the colour scale indicates the distribution of haloes. The eigenvalue of each PC is shown along the horizontal axis. 

PC1 clearly stands out as most important, accounting for no less than 37\% of the total variance, compared to 15\% for PC2. PCs 3 and 4 also account for $>10$\% of the variance, but the remaining PCs are not that significant. However, to account for 90\% of the total variance, we require no less than 6 PCs, a suprisingly large number. Even the least important PC still accounts for more than 2\% of the variance. Clearly, dark matter haloes are complicated objects whose structure cannot be described using a small number of parameters.

PC1 correlates very strongly with concentration ($\rank=-0.86$), which indicates that this parameter is most fundamental. Environment, spin, and particularly triaxiality also correlate significantly with PC1, but they correlate more strongly with other PCs. All other parameters correlate most strongly with PC1. This is consistent with the fact that its eigenvalue is by far the greatest. This confirms what Figure~\ref{ccdp1} suggested: of the parameters that we consider, concentration is most fundamental. 

PC2 correlates very strongly with spin ($\rank=0.72$), strongly with mass ($\rank=-0.56$), but not at all with concentration. PC3 only correlates strongly with environment ($\rank=0.68$). PC4 correlates most strongly with triaxiality ($\rank=0.64$), but is also strongly correlated with environment ($\rank=-0.55$). 

Hence, concentration, age, relaxedness, mass, and substructure all belong to a single family of parameters, which however, still contains a large amount of scatter. PCs 2-4 together account for the variance that is not linked with concentration. That part of the variance is mostly due to the scatter in spin, environment, and triaxiality, the three parameters that are most independent of the main family of halo properties.

As this paper was in the final stages of preparation, \cite{Skibba2011} posted a preprint of an independent, related study. While they also carried out a PCA, albeit with a somewhat different set of parameters, they did not present a correlation study. In agreement with our PCA results, they find that concentration is more fundamental than mass. However, they also find that relaxedness is about as important as concentration, whereas we find that the variation with relaxedness that does not trace concentration only features strongly in PC5, which accounts for only 5\% of the total variance. This difference may be due to the fact that their PCA includes multiple measures of relaxedness and that they did not consider age, which is closely related to concentration. Another cause of discrepancy is the use of different definitions for environment. They use the overdensity in a fixed aperture (of 8 Mpc$/h$), which  has been shown to correlate very strongly with mass \citep{Haas2011}, whereas our environmental parameter is insensitive to mass. 

To conclude, both the correlation analysis and the PCA demonstrate that concentration, age, substructure, mass, sphericity, and relaxedness are closely related, with concentration being most fundamental. Triaxiality, spin, and (mass-independent) environment are more independent, although spin correlates strongly with substructure and both spin and triaxiality are substantially correlated with concentration. While the scale of a halo is set by its mass, all other properties are more closely related to concentration.

\section*{Acknowledgements}
We thank Alan Duffy for providing us with the halo concentrations
and Volker Springel for allowing us to use \textsc{subfind} and \textsc{gadget}-3. The simulations were run on
the Cosmology Machine at the ICC in
Durham as part of the Virgo Consortium research programme. This work was supported by a Marie Curie reintegration grant and by the Initial Training Network CosmoComp.


\begin{thebibliography}{MBNB03}

\bibitem[\protect\citeauthoryear{Bett et al.}{2007}]{Bett2007} 
Bett P., Eke V., Frenk C.~S., Jenkins A., Helly J., Navarro J., 2007, 
MNRAS, 376, 215 

\bibitem[\protect\citeauthoryear{Bullock et 
al.}{2001}]{Bullock2001} Bullock J.~S., Dekel A., Kolatt T.~S., 
Kravtsov A.~V., Klypin A.~A., Porciani C., Primack J.~R., 2001, ApJ, 555, 
240 

\bibitem[\protect\citeauthoryear{Cooray 
\& Sheth}{2002}]{Cooray2002} Cooray~A., \& Sheth~R.\ 2002, Physics Reports, 372, 1 

\bibitem[\protect\citeauthoryear{Croton, Gao, 
\& White}{2007}]{Croton2007} Croton D.~J., Gao L., White S.~D.~M., 2007, MNRAS, 374, 1303 

\bibitem[\protect\citeauthoryear{Duffy et al.}{2008}]{Duffy2008} 
Duffy A.~R., Schaye J., Kay S.~T., Dalla Vecchia C., 2008, MNRAS, 390, L64 

\bibitem[\protect\citeauthoryear{Gao et al.}{2004}]{Gao2004} 
Gao L., White S.~D.~M., Jenkins A., Stoehr F., Springel V., 2004, MNRAS, 
355, 819 

\bibitem[\protect\citeauthoryear{Gao, Springel, 
\& White}{2005}]{Gao2005} Gao L., Springel V., White S.~D.~M., 2005, MNRAS, 363, L66 

\bibitem[\protect\citeauthoryear{Gao 
\& White}{2007}]{Gao2007} Gao L., White S.~D.~M., 2007, MNRAS, 377, L5 

\bibitem[\protect\citeauthoryear{Haas et al.}{2011}]{Haas2011} 
Haas M.~R., Schaye J., Jeeson-Daniel A., 2011, MNRAS, submitted, arxiv:1103.0547

\bibitem[\protect\citeauthoryear{Hahn et al.}{2007}]{Hahn2007} 
Hahn O., Porciani C., Carollo C.~M., Dekel A., 2007, MNRAS, 375, 489 

\bibitem[\protect\citeauthoryear{Harker et al.}{2006}]{Harker2006} 
 Harker G., Cole S., Helly J., Frenk C., Jenkins A., 2006, MNRAS, 367, 1039 

\bibitem[\protect\citeauthoryear{Jing, Suto, 
\& Mo}{2007}]{Jing2007} Jing Y.~P., Suto Y., Mo H.~J., 2007, ApJ, 657, 664 

\bibitem[\protect\citeauthoryear{Komatsu et 
al.}{2009}]{Komatsu2011} Komatsu E., et al., 2011, ApJS, 192, 18 

\bibitem[\protect\citeauthoryear{Li, Mo, 
\& Gao}{2008}]{Li2008} Li Y., Mo H.~J., Gao L., 2008, MNRAS, 389, 1419 

\bibitem[\protect\citeauthoryear{Macci{\`o} et 
al.}{2007}]{Maccio2007} Macci{\`o} A.~V., Dutton A.~A., van den 
Bosch F.~C., Moore B., Potter D., Stadel J., 2007, MNRAS, 378, 55 

\bibitem[\protect\citeauthoryear{Mo 
\& White}{1996}]{Mo1996} Mo H.~J., White S.~D.~M., 1996, MNRAS, 282, 347 

\bibitem[\protect\citeauthoryear{Navarro, Frenk, 
\& White}{1997}]{Navarro1997} Navarro J.~F., Frenk C.~S., White S.~D.~M., 1997, ApJ, 490, 493 

\bibitem[\protect\citeauthoryear{Seljak \& Zaldarriaga}{1996}]{Seljak1996} Seljak, U., \& Zaldarriaga, M.\ 1996, ApJ, 469, 437 

\bibitem[\protect\citeauthoryear{Skibba 
\& Macci{\`o}}{2011}]{Skibba2011} Skibba R.~A., Macci{\`o} A.~V., 2011, arXiv, arXiv:1103.1641 

\bibitem[\protect\citeauthoryear{Springel et 
al.}{2001}]{Springel2001} Springel V., White S.~D.~M., Tormen G., 
Kauffmann G., 2001, MNRAS, 328, 726 

\bibitem[\protect\citeauthoryear{Springel}{2005}]{Springel2005} 
Springel V., 2005, MNRAS, 364, 1105

\bibitem[\protect\citeauthoryear{Taylor}{2011}]{Taylor2011} Taylor 
J.~E., 2011, AdAst, 2011, 6 

\bibitem[\protect\citeauthoryear{Wang et al.}{2010}]{Wang2010} 
Wang H., Mo H.~J., Jing Y.~P., Yang X., Wang Y., 2010, arXiv, 
arXiv:1007.0612 

\bibitem[\protect\citeauthoryear{Wechsler et 
al.}{2006}]{Wechsler2006} Wechsler R.~H., Zentner A.~R., Bullock 
J.~S., Kravtsov A.~V., Allgood B., 2006, ApJ, 652, 71 

\bibitem[\protect\citeauthoryear{Wetzel et al.}{2007}]{Wetzel2007} 
 Wetzel A.~R., Cohn J.~D., White M., Holz D.~E., Warren M.~S., 2007, ApJ, 656, 139 

\bibitem[\protect\citeauthoryear{Zhao et al.}{2003}]{Zhao2003} 
Zhao D.~H., Jing Y.~P., Mo H.~J., B{\"o}rner G., 2003, ApJ, 597, L9 

\end{thebibliography}
\end{document}